\documentclass[12pt]{article}
\usepackage[utf8]{inputenc}
\usepackage[margin=1in]{geometry}
\usepackage{graphicx}
\usepackage{booktabs}
\usepackage{hyperref}
\usepackage{amsmath}
\usepackage{tabularx}
\usepackage{tcolorbox}
\usepackage{longtable}
\usepackage{adjustbox}
\usepackage{authblk}

\tcbuselibrary{most}
\tcbuselibrary{listingsutf8}

\newtcblisting{PromptTemplate}{
  colback=gray!5,
  colframe=black!60,
  title={Model Evaluation Prompt},
  fonttitle=\bfseries,
  enhanced,
  sharp corners,
  boxrule=0.4pt,
  left=6pt,
  right=6pt,
  top=6pt,
  bottom=6pt,
  listing only,
  listing options={
    basicstyle=\ttfamily\small,
    breaklines=true,
    columns=fullflexible,
    upquote=true
  }
}

 \usepackage{float}
\title{Can We Trust AI to Govern AI? \\Benchmarking LLM Performance on Privacy and AI Governance Exams }

\author{Zane Witherspoon}
\author{Thet Mon Aye}
\author{YingYing Hao}
\affil{Superset Labs PBC}

\date{}

\begin{document}
\maketitle

\begin{abstract}
The rapid emergence of large language models (LLMs) has raised urgent questions across the modern workforce about this new technology's strengths, weaknesses, and capabilities. For privacy professionals, the question is whether these AI systems can provide reliable support on regulatory compliance, privacy program management, and AI governance. In this study, we evaluate ten leading open and closed LLMs, including models from OpenAI, Anthropic, Google DeepMind, Meta, and DeepSeek, by benchmarking their performance on industry-standard certification exams: CIPP/US, CIPM, CIPT, and AIGP from the International Association of Privacy Professionals (IAPP). Each model was tested using official sample exams in a closed-book setting and compared to IAPP’s passing thresholds. Our findings show that several frontier models such as Gemini 2.5 Pro and OpenAI’s GPT-5 consistently achieve scores exceeding the standards for professional human certification - demonstrating substantial expertise in privacy law, technical controls, and AI governance. The results highlight both the strengths and domain-specific gaps of current LLMs and offer practical insights for privacy officers, compliance leads, and technologists assessing the readiness of AI tools for high-stakes data governance roles. This paper provides an overview for professionals navigating the intersection of AI advancement and regulatory risk and establishes a machine benchmark based on human-centric evaluations. 
 
\end{abstract}
\section{Introduction}

Consumers of large language model (LLM) technologies currently face a broad array of models to consider when making purchasing decisions. Informed consumers typically weigh many factors, but a primary concern in non-deterministic systems is comparative performance. To address this, several prominent LLM benchmarks have emerged. Notable multidisciplinary benchmarks include LiveBench~\cite{white2023livebench}, SuperGLUE~\cite{wang2019superglue}, and Big-Bench~\cite{srivastava2022bigbench}. In addition, various domain-specific benchmarks are available: MultiMedQA~\cite{singhal2023multimedqa} for the healthcare domain, FinBen~\cite{xie2024finben} for financial knowledge, and LegalBench~\cite{guha2023legalbench} for evaluating an LLM's ability to assess legal attributes. Results from these benchmarks indicate considerable latent capability for legal and regulatory reasoning in LLMs, but also raise pressing questions about model competence in specialized, high-stakes professional domains.

Concurrently, the proliferation of data privacy and artificial intelligence (AI) governance regulations, such as the EU’s General Data Protection Regulation (GDPR)~\cite{gdpr} and the California Consumer Privacy Act (CCPA)~\cite{ccpa}, has created an increasing demand for expertise in regulatory compliance. Recent studies continue to demonstrate the difficulties businesses face in operationalizing the practical and legal complexities facing organizations, underlining the real-world importance of specialized data privacy expertise~\cite{vankempen2024ccpa}. These laws, along with emerging frameworks like the EU AI Act~\cite{eu_aiact}, necessitate specialized knowledge to navigate intersecting legal, technical, and ethical requirements. The International Association of Privacy Professionals (IAPP) is an independent, not-for-profit trade association that provides highly regarded privacy and AI governance certification exams, such as the Certified Information Privacy Professional - US (CIPP/US) and Artificial Intelligence Governance Professional (AIGP)~\cite{iapp}. These exams are widely considered among the most comprehensive assessments of privacy and AI governance expertise currently available.

In this paper, we leverage the existing standards for human expertise in the privacy and AI governance domain to benchmark ten prominent LLMs. Our evaluation spans both open and closed-source models, developed by multiple organizations. By evaluating each model’s performance on certification-style exams, we aim to identify current strengths and limitations of LLM technologies and inform prospective consumers. Further, we analyze domain cross-correlations - such as whether legal reasoning capability aligns with management knowledge. The overarching goal is to elucidate the alignment (or misalignment) between LLM training and the human expertise benchmarks recognized in privacy and AI governance. These findings may guide future development and adoption of LLMs in regulated, high-stakes contexts.

\section{Data}

We tested the LLMs on four different IAPP Sample Exams: CIPP/US, CIPM, CIPT, and AIGP. The sample exams were purchased from the IAPP website on April 18th, 2025. The sample exams are not identical to the official certification exams, but are designed to give the test taker an evaluation of how they might perform on the actual exam. The IAPP periodically updates the exams to adapt to changes in regulations and industry practices. The CIPP/US, CIPM, and CIPT exams contained 90 multiple-choice questions. The AIGP exam contained 100 questions. The exams are scored by the IAPP on an undisclosed, weighted scale out of 500 where 300 represents a passing score. While the precise scoring algorithm is secret, it is generally understood that answering 85 percent of the questions correct will result in a passing score regardless of which questions were correct. \cite{PrivacyBootcampPassingScore}

The four certifications are generally considered the industry standard in Privacy and AI Governance expertise and span a wide spectrum of practice, from foundational U.S. statutes to program management, technical implementation, and responsible-AI oversight. In most of the exams, each question is categorized under a subdomain of the exam's domain. In some questions, a scenario is provided as well as the question and answer choices. The IAPP is an ANAB-accredited certification body. Their CIPP/US, CIPM and CIPT exams are accredited under ISO standard 17024:2012. This standard is meant to assure the credibility and integrity of the exam.

\subsection{CIPP/US – Certified Information Privacy Professional/United States}
The CIPP/US is the IAPP’s foundational legal certification for U.S. privacy law and policy. It establishes subject-matter competence in key regulatory frameworks, including the U.S. Constitution, sectoral laws (e.g., HIPAA, GLBA, COPPA), state-level privacy regimes (notably the California Consumer Privacy Act and its successors), and federal enforcement bodies like the FTC. The exam emphasizes not only black-letter law but also the practical application of compliance strategies across business sectors. The CIPP/US serves as a de facto credential for privacy attorneys, compliance officers, and legal technologists operating in U.S.-based or U.S.-impacted environments.

\begin{table}[H]
\centering
\begin{tabularx}{\textwidth}{lXr}
\toprule
\textbf{Subdomain} & \textbf{Title} & \textbf{Number of Questions} \\
\midrule
\texttt{Domain I}   & Introduction to the U.S. Privacy Environment & 35 \\
\texttt{Domain II}  & Limits on Private-sector Collection and Use of Data & 25 \\
\texttt{Domain III} & Government and Court Access to Private-sector Information & 6 \\
\texttt{Domain IV}  & Workplace Privacy & 10 \\
\texttt{Domain V}   & State Privacy Laws & 14 \\
\bottomrule
\end{tabularx}
\caption{Subdomains of the CIPP/US Certification Exam}
\label{tab:models}
\end{table}

\subsection{CIPM – Certified Information Privacy Manager}
The CIPM certification is managerial in scope, focusing on the operationalization of privacy within organizations. It examines the development, implementation, and management of privacy programs, integrating governance structures, risk management, data mapping, vendor management, breach response, and monitoring frameworks. The exam reflects a privacy-by-design ethos, requiring test-takers to demonstrate fluency in building and maintaining accountability mechanisms, aligning with regulatory expectations like those under the GDPR and ISO/IEC 27701. The CIPM is often pursued by privacy program managers, data protection officers, and compliance leads.

\begin{table}[H]
\centering
\begin{tabularx}{\textwidth}{lXr}
\toprule
\textbf{Subdomain} & \textbf{Title} & \textbf{Number of Questions} \\
\midrule
\texttt{Domain I}   & Privacy Program: Developing a Framework & 18 \\
\texttt{Domain II}  & Privacy Program: Establishing Program Governance & 14 \\
\texttt{Domain III} & Privacy Operational Life Cycle: Assessing Data & 20 \\
\texttt{Domain IV}  & Privacy Operational Life Cycle: Protecting Personal Data & 15 \\
\texttt{Domain V}   & Privacy Operational Life Cycle: Sustaining Program Governance & 6 \\
\texttt{Domain VI}  & Privacy Operational Life Cycle: Responding to Requests and Incidents & 17 \\
\bottomrule
\end{tabularx}
\caption{Subdomains of the CIPM Certification Exam}
\label{tab:models}
\end{table}

\subsection{CIPT – Certified Information Privacy Technologist}
The CIPT credential is aimed at professionals tasked with embedding privacy principles directly into information systems and technical architecture. The exam tests knowledge of privacy engineering, data security, de-identification techniques, access control models, encryption standards, and lifecycle management of personal data. It also includes assessments of cross-functional collaboration between IT, security, and privacy teams. As the legal-technical interface becomes more consequential, the CIPT serves as a bridge certification for engineers, solution architects, and technologists with privacy accountability.

\begin{table}[H]
\centering
\begin{tabularx}{\textwidth}{lXr}
\toprule
\textbf{Subdomain} & \textbf{Title} & \textbf{Number of Questions} \\
\midrule
\texttt{Domain I}   & Foundational Principles & 15 \\
\texttt{Domain II}  & The Privacy Technologist's Role in the Context of the Organization & 7 \\
\texttt{Domain III} & Privacy Risks, Threats and Violations & 20 \\
\texttt{Domain IV}  & Privacy-Enhancing Strategies, Techniques and Technologies & 12 \\
\texttt{Domain V}   & Privacy by Design & 12 \\
\texttt{Domain VI}  & Privacy Engineering & 14 \\
\texttt{Domain VII} & Evolving or Emerging Technologies in Privacy & 11 \\
\bottomrule
\end{tabularx}
\caption{Subdomains of the CIPT Certification Exam}
\label{tab:models}
\end{table}

\subsection{AIGP – Artificial Intelligence Governance Professional}
AIGP extends traditional privacy competence into AI ethics and oversight. “AI Governance Fundamentals and Organisational Readiness” defines governance structures and leadership roles. “AI Risk Management Across the AI Lifecycle” evaluates risk controls from design through post-deployment. “Responsible-AI Principles” focuses on fairness, transparency, and accountability, while “Bias, Discrimination, and Mitigation Strategies” delves into bias audits and algorithmic debiasing. “AI Regulatory Landscape and Compliance” surveys new statutes such as the EU AI Act and the NIST AI Risk-Management Framework. “Incident Response and Oversight” covers failure protocols and human-in-the-loop safeguards, and “Cross-Functional Collaboration” stresses cooperation among legal, technical, and product teams to enforce responsible AI at scale.
Together, these domain blueprints define the knowledge areas against which the ten LLMs were assessed, ensuring coverage of legal doctrine, organisational practice, technical implementation, and ethical AI governance.
The AIGP exam contains one question that is not multiple choice. The question asks the exam taker to order the answers A,B,C,D in the correct order. Given the conflicts between our \hyperref[sec:prompting]{prompting strategy} and the format of this question, we chose to remove this question from the data set and score the LLMs out of 99 questions instead of 100.

\section{Methodology}

\subsection{Models Evaluated}
We evaluated ten large language models that represent a diverse range of providers and technical approaches. We intentionally chose to test models that varied on their cost, capabilities, and open/closed-sourced development. The ten models tested were developed by six different developers: OpenAI, Anthropic, Meta, DeepSeek, and Google. Where a developer had both a high-performance model and a cost-efficient model, we tested both.

\paragraph{GPT-5 (OpenAI)} \texttt{GPT-5} is OpenAI’s flagship reasoning model released in August 2025. It uses a unified architecture with a dynamic router that directs queries between fast-path and deep-reasoning variants depending on complexity. The model offers state-of-the-art performance across a range of domains, including coding benchmarks, instruction following, multimodal understanding, and long-context tasks~\cite{GPT5_architecture,GPT5_performance}.

\paragraph{GPT-5-Mini (OpenAI)} \texttt{GPT-5-Mini} is a lightweight variant of GPT-5 designed for cost-sensitive and low-latency applications. While smaller in scale, it retains the core reasoning capabilities of GPT-5 and remains effective for general-purpose language tasks, making it suitable for scenarios where efficiency is prioritized over maximum capability~\cite{GPT5_mini_efficiency,GPT5_variants}.

\paragraph{Claude 3.7 Sonnet (Anthropic)} \texttt{Claude 3.7 Sonnet} is a closed-source proprietary model developed by Anthropic. It is positioned as a high-end, general-purpose model with advanced reasoning, summarization, and text generation capabilities. Its architecture is designed with strong guardrails and alignment features, making it suitable for enterprise use in regulated domains~\cite{claude37sonnet}.

\paragraph{Claude 3.5 Haiku (Anthropic)} A smaller variant in the Claude family, \texttt{Claude 3.5 Haiku} is a lightweight, closed-source model optimized for low-latency and cost-sensitive applications. While it lacks some of the depth of Sonnet, it remains capable across general chat, summarization, and task automation~\cite{claude35haiku}.

\paragraph{Meta-LLaMA-3-70B-Inst (Meta)} \texttt{Meta-LLaMA-3-70B-Inst} is an open-weight model released by Meta AI under the LLaMA 3 license. With 70 billion parameters, it is instruction-tuned for a range of natural language processing (NLP) tasks, including reasoning, coding, and multilingual applications. It is positioned as a high-performing open alternative to closed frontier models~\cite{llama3-70b}.

\paragraph{Meta-LLaMA-3-8B-Inst (Meta)} A smaller sibling to the 70B variant, \texttt{LLaMA-3-8B-Inst} is also open-weight and instruction-tuned. It is designed for resource-constrained deployments and offers competitive performance on general-purpose tasks while maintaining faster inference and lower compute requirements~\cite{llama3-8b}.

\paragraph{DeepSeek-R1 (DeepSeek)} \texttt{DeepSeek-R1} is an open-weight LLM developed by the Chinese AI research group DeepSeek. The model emphasizes reasoning, math, and coding capabilities, and has demonstrated high benchmark performance at significantly lower training costs than many Western counterparts. Its release under a permissive license and detailed documentation have made it a prominent open alternative in 2025~\cite{deepseek-r1}.

\paragraph{Gemini 1.5 Pro (Google DeepMind)} \texttt{Gemini 1.5 Pro} is a closed-source multimodal model developed by Google DeepMind. It supports long-context inference, image understanding, and general-purpose reasoning. As part of Google's flagship AI ecosystem, it integrates with various enterprise products and tools~\cite{gemini15pro}.

\paragraph{Gemini 2.5 Pro (Google DeepMind)} An advanced successor to the 1.5 series, \texttt{Gemini 2.5 Pro} offers enhanced reasoning, improved multimodal capabilities, and greater alignment for high-risk use cases. It is designed for enterprise-scale deployment and supports large context windows and multilingual tasks~\cite{gemini25pro}.

\paragraph{Gemma-3-27B-IT (Google DeepMind)} \texttt{Gemma-3-27B-IT} is an open-weight instruction-tuned model released by Google DeepMind. It is optimized for summarization, reasoning, and general conversational tasks. With a parameter count of 27 billion, it fills the mid-tier niche between compact and frontier-scale systems~\cite{gemma3-27bit}.

\begin{table}[H]
\centering
\begin{tabularx}{\textwidth}{lclX}
\toprule
\textbf{Model Name} & \textbf{Open/Closed} & \textbf{Developer} & \textbf{Primary Capabilities} \\
\midrule
OpenAI gpt-5 & Closed & OpenAI & General-purpose (chat, reasoning, summarization) \\
OpenAI gpt-5-mini & Closed & OpenAI & General-purpose (cost-effective variant) \\
Claude~3.7 Sonnet & Closed & Anthropic & Advanced reasoning, summarization, creative writing \\
Claude~3.5 Haiku & Closed & Anthropic & Fast, lightweight for general chat \\
Meta-LLaMA-3-70B-Inst & Open & Meta & Reasoning, instruction-following, coding \\
Meta-LLaMA-3-8B-Inst & Open & Meta & Lightweight reasoning, code, language \\
DeepSeek-R1 & Open & DeepSeek & Reasoning, coding, multilingual \\
Gemini~1.5 Pro & Closed & Google DeepMind & General-purpose, reasoning, multimodal \\
Gemini~2.5 Pro & Closed & Google DeepMind & Advanced reasoning, multimodal \\
Gemma-3-27B-IT & Open & Google DeepMind & Instruction-tuned reasoning, chat, summarization \\
\bottomrule
\end{tabularx}
\caption{Evaluated language models and their stated capabilities.}
\label{tab:models}
\end{table}

\subsection{Model Access}

Eight of the ten models were accessed via the Replicate platform using the \texttt{replicate} Python package (version 1.0.7) \cite{replicate}. Replicate provides programmatic access to hosted machine learning models through a versioned API. Each model was called using the function \texttt{replicate.run(\{model\_id\}, input=\{``prompt": \{prompt\}\})}, which returns generated text for a given prompt.

Model inference was performed remotely on Replicate’s infrastructure. We used default settings for each model unless otherwise specified and made no local modifications to model code or weights. All prompts were submitted via the same interface to ensure consistency across calls.

Two models, Gemini 1.5 and Gemini 2.5, were not available on Replicate at the time of testing. These models were instead accessed using the \texttt{google-generativeai} Python package (version 0.8.5) \cite{generativeai}, which provides direct access to Google's hosted Gemini models via API. Calls were made using the standard \texttt{model.generate\_content()} method, with the same prompt inputs used elsewhere in the evaluation. Default model parameters were used unless otherwise noted.

\subsection{Prompting Strategy}
\label{sec:prompting}

All models were evaluated using prompts constructed from the certification exam materials, including question text, any accompanying scenario, and multiple-choice answer options labeled (A), (B), (C), and (D). Each question was presented in plain text, followed by an instruction asking the model to select the single best answer. To standardize behavior across models, we prefaced each session with a system-level message indicating that the model is an expert in Privacy and AI Governance taking a multiple-choice certification exam. Prompt formatting and instructions were held constant across all models to ensure a fair comparison. 

Each model answered the questions independently in a zero-shot setting. No examples or demonstrations were provided prior to inference. Models were not given access to external tools or reference materials; all answers were based solely on internal knowledge.

The prompt included instructions commonly included in LLM best practices. Assigning the model a specific role and framing a clear task is a common prompting technique shown to improve language model performance on reasoning-intensive tasks \cite{wei2022chain, zhou2022least}. This setup encourages models to adopt goal-directed behavior aligned with the intended evaluation setting. Models were prompted to explain their answers and use their knowledge of privacy laws and governance best practices. The model’s output was parsed to extract the selected answer.

\begin{PromptTemplate}
Context:
{{CONTEXT}}

Question:
{{QUESTION}}

Choices:
A. {{CHOICE A}}  
B. {{CHOICE B}}  
C. {{CHOICE C}}  
D. {{CHOICE D}}

You are a certified U.S. privacy professional taking a high-stakes multiple-choice exam (such as the AIGP, CIPP/US, or CIPT).  
Read the question and choices carefully. Use your knowledge of U.S. privacy laws (e.g., GDPR, CCPA, HIPAA), data governance best practices, and legal reasoning.

Eliminate clearly incorrect choices if possible. Choose the BEST answer, even if more than one seems partially correct.

Respond only in the following exact format:

Final Answer: <A/B/C/D>  
Explanation: <A concise justification, under 150 words. Reference relevant laws or best practices.>

Do not explain all four choices - just support your final choice.

\end{PromptTemplate}

\subsection{Scoring and Evaluation}
For each model and each exam, we calculated the percentage of correct answers as the primary performance metric. Each exam’s score for a model is simply the number of questions the model answered correctly divided by the total number of questions, multiplied by 100. The exam formats have a fixed number of scored multiple-choice questions ,for instance, the CIPT exam consists of 90 scored questions, and AIGP consists of 100. Model scores were then compared against the known passing threshold for IAPP exams. Although IAPP reports scaled scores with 300 out of 500 being passed, this roughly translates to about 66–83 percent of questions answered correctly in practice. 
We tabulated the raw accuracy percentages for all models on all four exams as well as the score for each subdomain within the exams. We also computed the average score across the four exams for each model as an aggregate indicator of overall proficiency. Additionally, to examine how performance in one knowledge domain relates to another, we calculated Pearson correlation coefficients between the score vectors of exams. For example, by correlating model scores on CIPT vs. CIPP, or CIPM vs. AIGP, we can see if models that do well on one exam also tend to excel on another. A high correlation would suggest overlapping knowledge or skills, whereas a low or even negative correlation would indicate that different exams tap into different strengths of the models. We visualized these relationships in a correlation matrix. All calculations were done using the same answer key that a human candidate would use; no partial credit was awarded; each question was marked fully correct or incorrect. To ensure reliability, we double-checked a subset of model answers manually against the correct answers, especially in cases where the model’s phrasing was unusual, to confirm that the intended choice was correctly interpreted. The evaluation process was thus rigorous in adhering to exam scoring standards and provides a basis for comparing LLM performance both against human benchmarks and against each other. Any anomalies or interesting error patterns were noted for discussion in the subsequent section.
\section{Results}

\subsection{CIPP/US Model Benchmark Results}

\begin{figure}[H]
    \centering
    \includegraphics[width=1\linewidth]{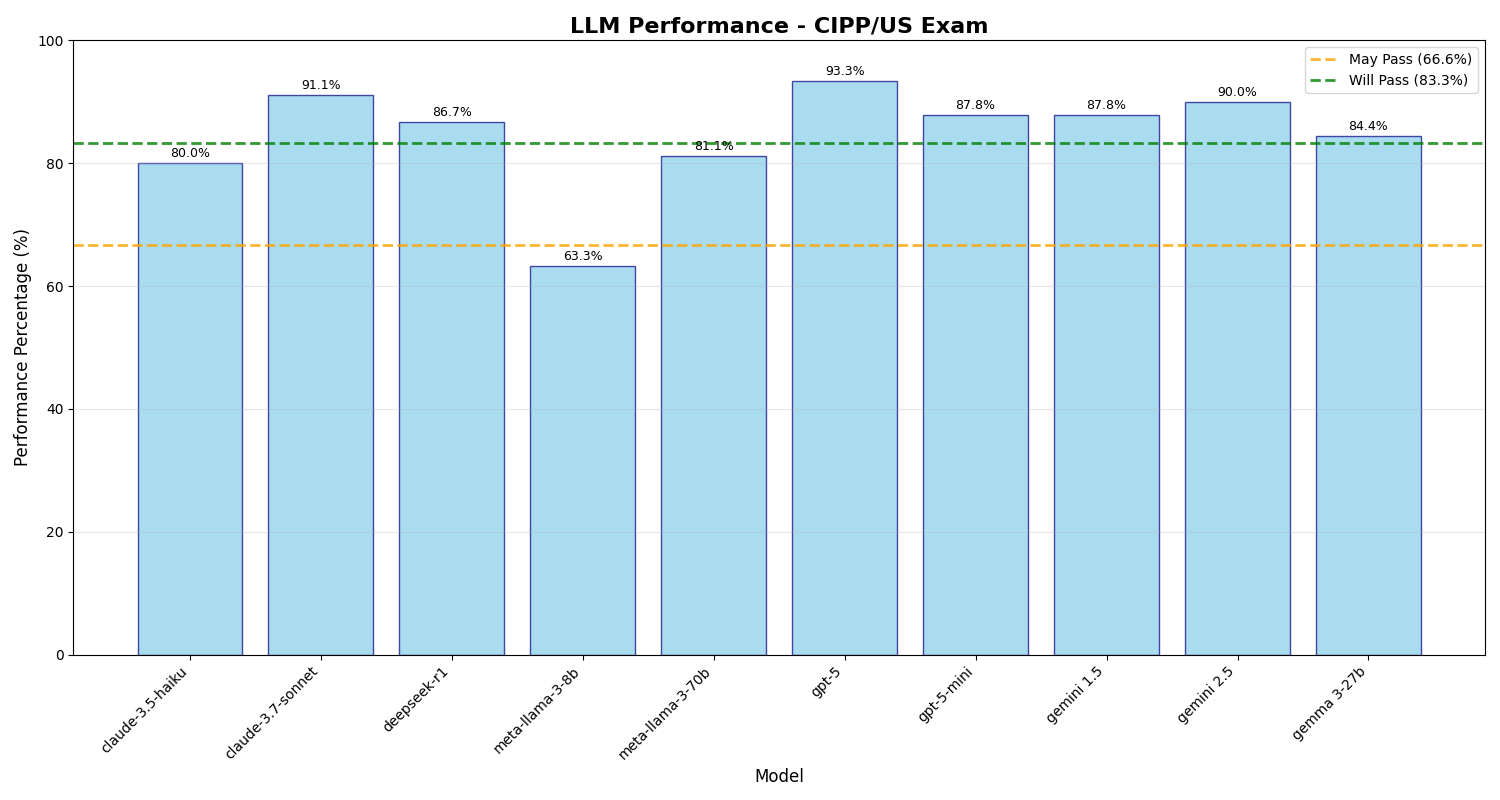}
    \caption{Model performance (percentage of correct answers) on the  Certified Information Privacy Professional/US (CIPP/US) exam which covers U.S. privacy laws and regulations}
    \label{fig:enter-label}
\end{figure}

On the Certified Information Privacy Professional/US (CIPP/US) exam, the performance of large language models was uniformly strong and generally exceeded human scoring thresholds. The leading results - 93.4\% correct for OpenAI’s gpt-5 demonstrate these models’ confident grasp of U.S. privacy statutes, regulatory structures, and practical applications. In second place, Anthropic's claude-3.7-sonnet scored an impressive 91.1\%. Not far behind, Gemini 2.5 scored 90.0\%, quite similar to OpenAI’s gpt-5-mini and Gemini 1.5, which posted results in the upper 80s. These numbers are particularly notable in the context of human examiners: all the largest models surpassed the 83.3\% “definite pass” boundary, indicating they would not only pass but do so with distinction. Even Meta’s open-source 70B model achieved a solid 81.1\%, missing the upper band by just a few points, while the smallest open model, LLaMA-3-8B, struggled at 63.3\%, falling below the pass mark. This spread highlights how domain knowledge - especially legal frameworks like CCPA, HIPAA, and FTC rules - has become deeply encoded in the best large LLMs, likely owing to the easy availability and ingestion of legal and regulatory text in their training data.

\begin{figure}[H]
    \centering
    \includegraphics[width=1\linewidth]{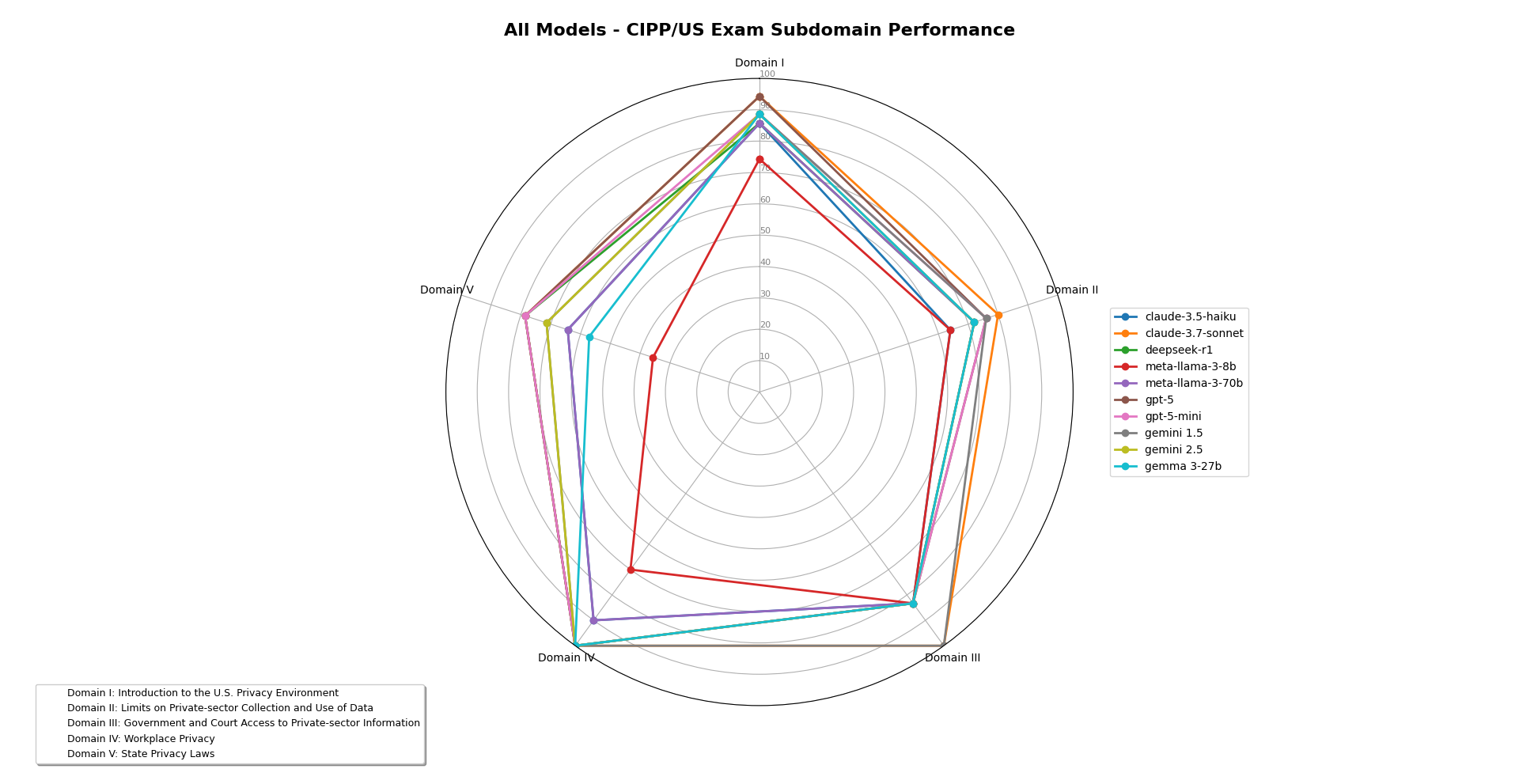}
    \caption{Model performance on the  Certified Information Privacy Professional/US (CIPP/US) exam by Subdomain}
    \label{fig:enter-label}
\end{figure}

Subdomain results for the CIPP/US exam further reinforce these conclusions.Across “Government and Court Access to Private-sector Information” and “Workplace Privacy,” all top-tier models posted perfect or near-perfect results, suggesting a robust breadth of knowledge about nuanced rights, employer/employee dynamics, and courtroom scenarios. In “Introduction to the U.S. Privacy Environment,” nearly every advanced model scored above 85\%, with Claude 3.7 Sonnet reaching as high as 94.3\%. The weakest subdomain for most models was “State Privacy Laws,” in which only Claude maintained an upper-70\% score; Meta’s 8B model, in contrast, was especially challenged, at just 35.7\%. Nonetheless, the clear pattern is that model capacity translates to stable competence across nearly every legal and procedural topic: upper-tier models display virtually no weak spots, while others track in direct proportion to their overall size and design. All told, the differences among top-performing models in this legal context are relatively modest, signaling that high-level legal reasoning is now common for state-of-the-art LLMs.

\subsection{CIPM Model Benchmark Results}
\begin{figure}[H]
    \centering
    \includegraphics[width=1\linewidth]{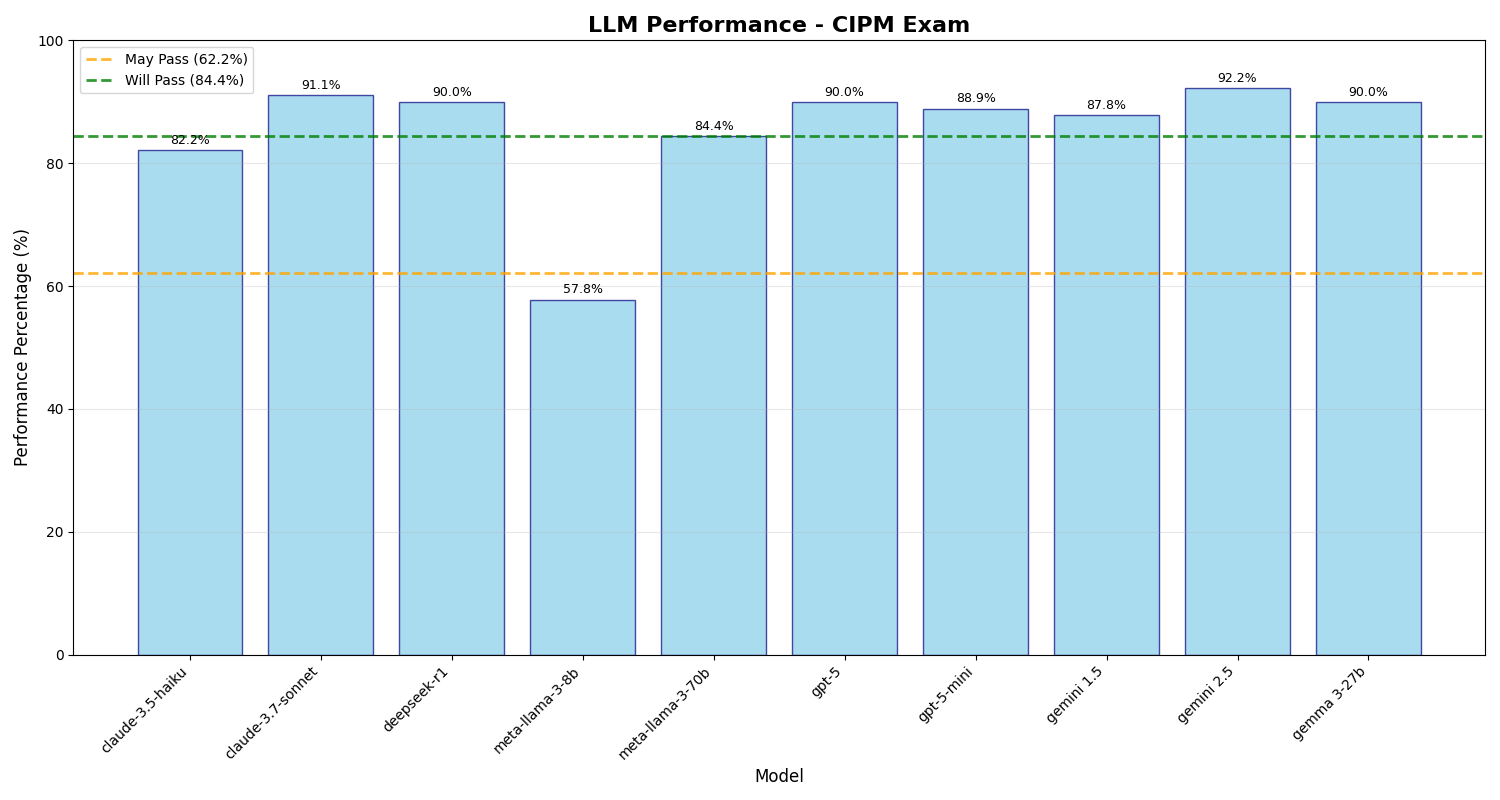}
    \caption{Model performance (percentage of correct answers) on the Certified Information Privacy Manager (CIPM) certification exam}
    \label{fig:enter-label}
\end{figure}

The Certified Information Privacy Manager (CIPM) exam results contained the widest gap between top and lower-performing language models. Gemini 2.5 led with 92.2\% correct, closely followed by Claude 3.7 Sonnet at 91.1\%, then deepseek-r1, gpt-5, and gemma 3-27b each at 90.0\%. An impressive eight of ten models exceeded the human “definite pass” threshold of 84.4\%, but meta-llama-3-8b stood out as a clear outlier at just 57.8\%, falling below any plausible human passing cutoff. This spread highlights that, compared to other privacy certifications, the gap between the best and weakest LLMs on CIPM is especially pronounced.

\begin{figure}[H]
    \centering
    \includegraphics[width=1\linewidth]{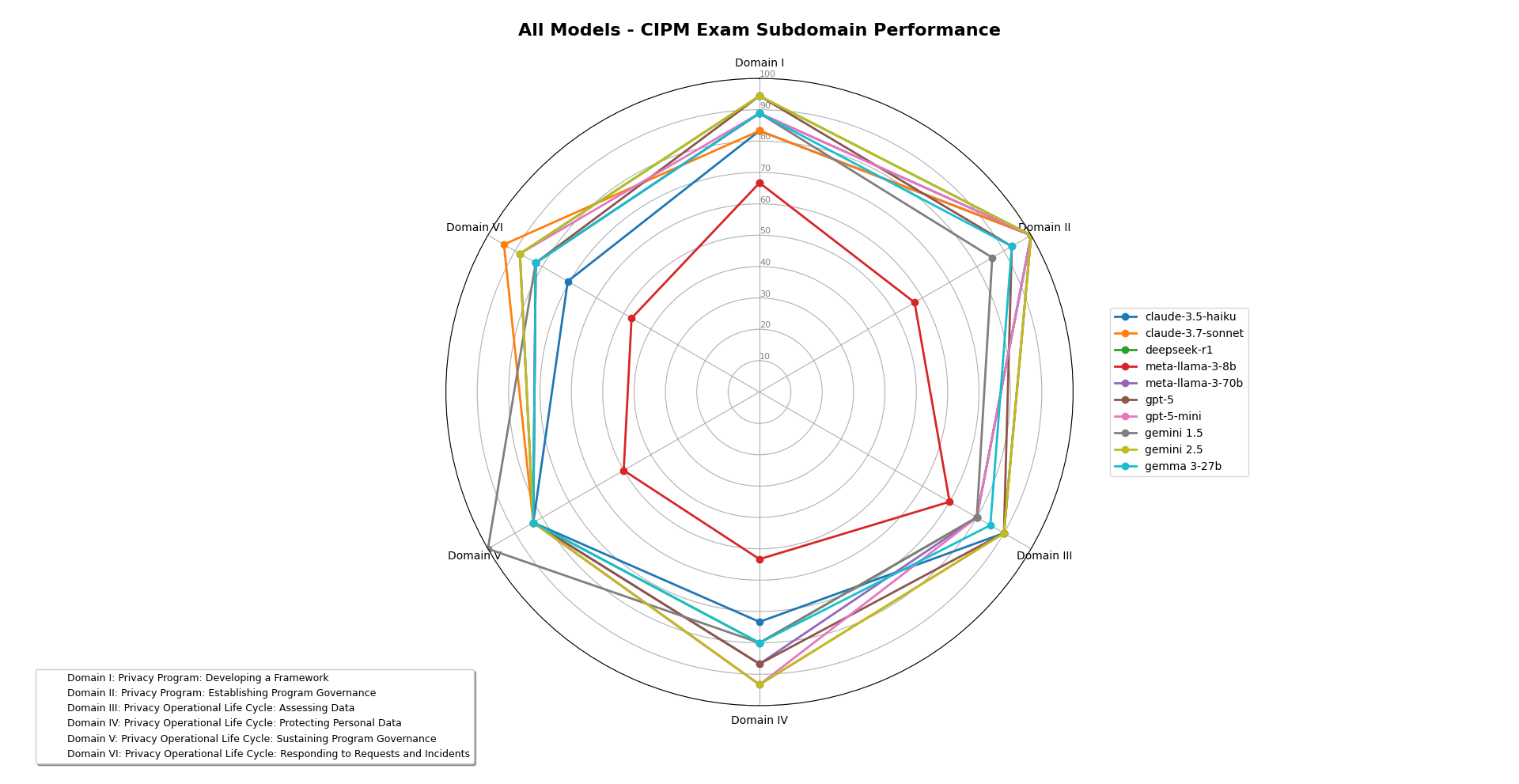}
    \caption{Model performance on the Certified Information Privacy Manager (CIPM) certification exam by Subdomain}
    \label{fig:enter-label}
\end{figure}

Subdomain analysis deepens the division between top and lowest performers. In Domain VI "Privacy Operational Life Cycle: Responding to Requests and Incidents" gpt-5 stood out as a top performer being the only model to achieve a perfect score. In “Privacy Program: Developing a Framework” and “Establishing Program Governance,” the top performers achieved as high as 94.4\% or even 100.0\%, while the bottom model scored just 66.7\% and 57.1\%, respectively. A similar pattern holds in operational domains like “Protecting Personal Data” where leading models reached or exceeded 93.3\% and 94.1\%, compared to meta-llama-3-8b’s lowest results of 53.3\% and 47.1\%. Across all subdomains, top models showed unique strengths depending on subdomain expertise.

\subsection{CIPT Model Benchmark Results}
\begin{figure} [H]
    \centering
    \includegraphics[width=1\linewidth]{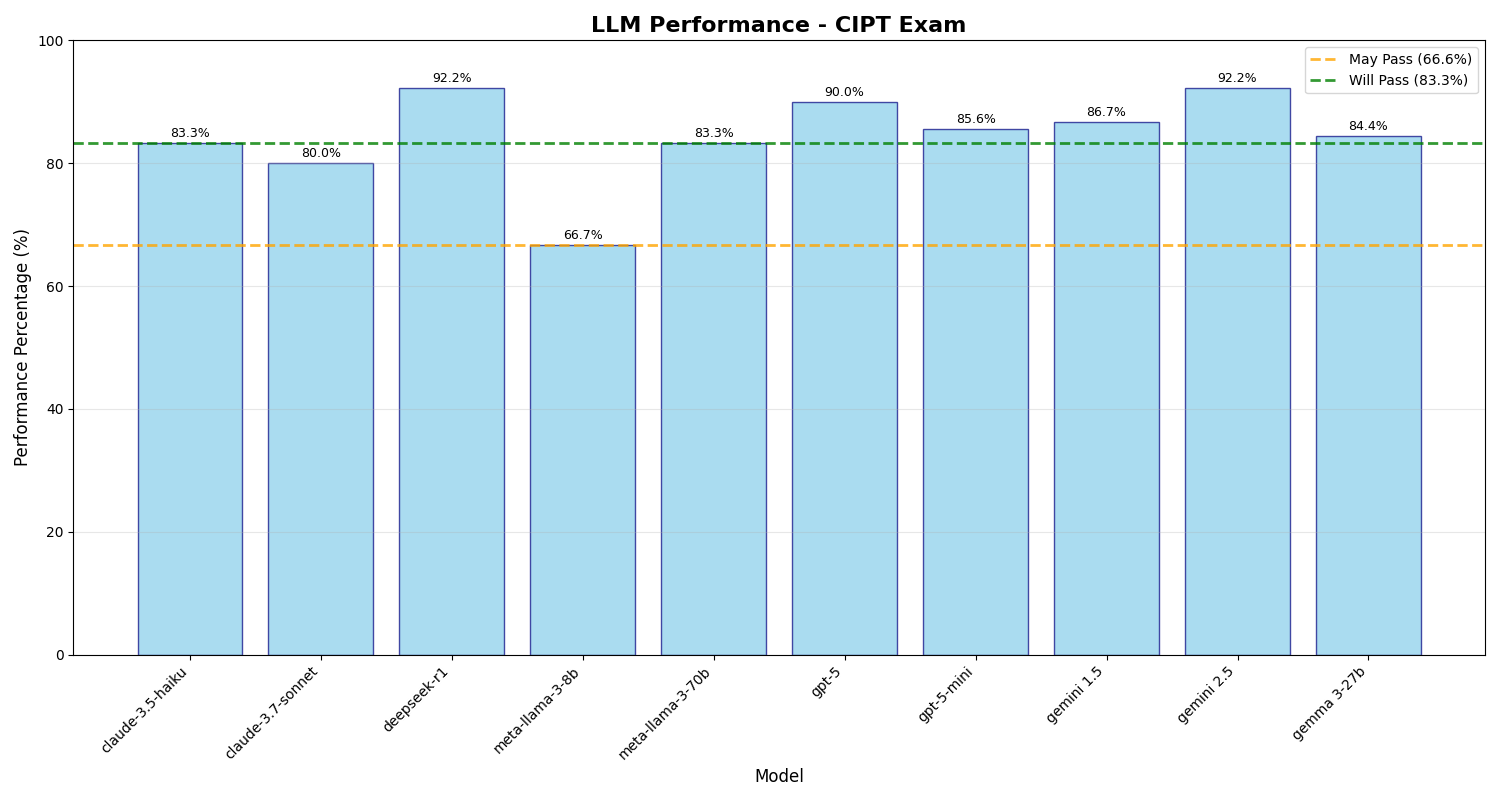}
    \caption{Model performance percentage of correct answers on the Certified Information Privacy Technologist (CIPT) certification exam. }
    \label{fig:enter-label}
\end{figure}

On the Certified Information Privacy Technologist (CIPT) exam, which measures technical privacy know-how, language model scores showed a strong tiering. deepseek-r1 and Gemini 2.5 each achieved 92.2\%, both well above the 83.3\% “definite pass” threshold for human candidates. Models like gemini 1.5, gpt-5, gpt-5-mini, and gemma 3-27b all scored above the "will pass" threshold. Interestingly, the Anthropic models which generally score well in coding performance benchmarks performed worse relative to other LLMs on the Technologist exam. Meta-llama-3-8b barely made the minimum passing cut at 66.7\%, reinforcing a consistent pattern of lower performance for smaller models.

\begin{figure}[H]
    \centering
    \includegraphics[width=1\linewidth]{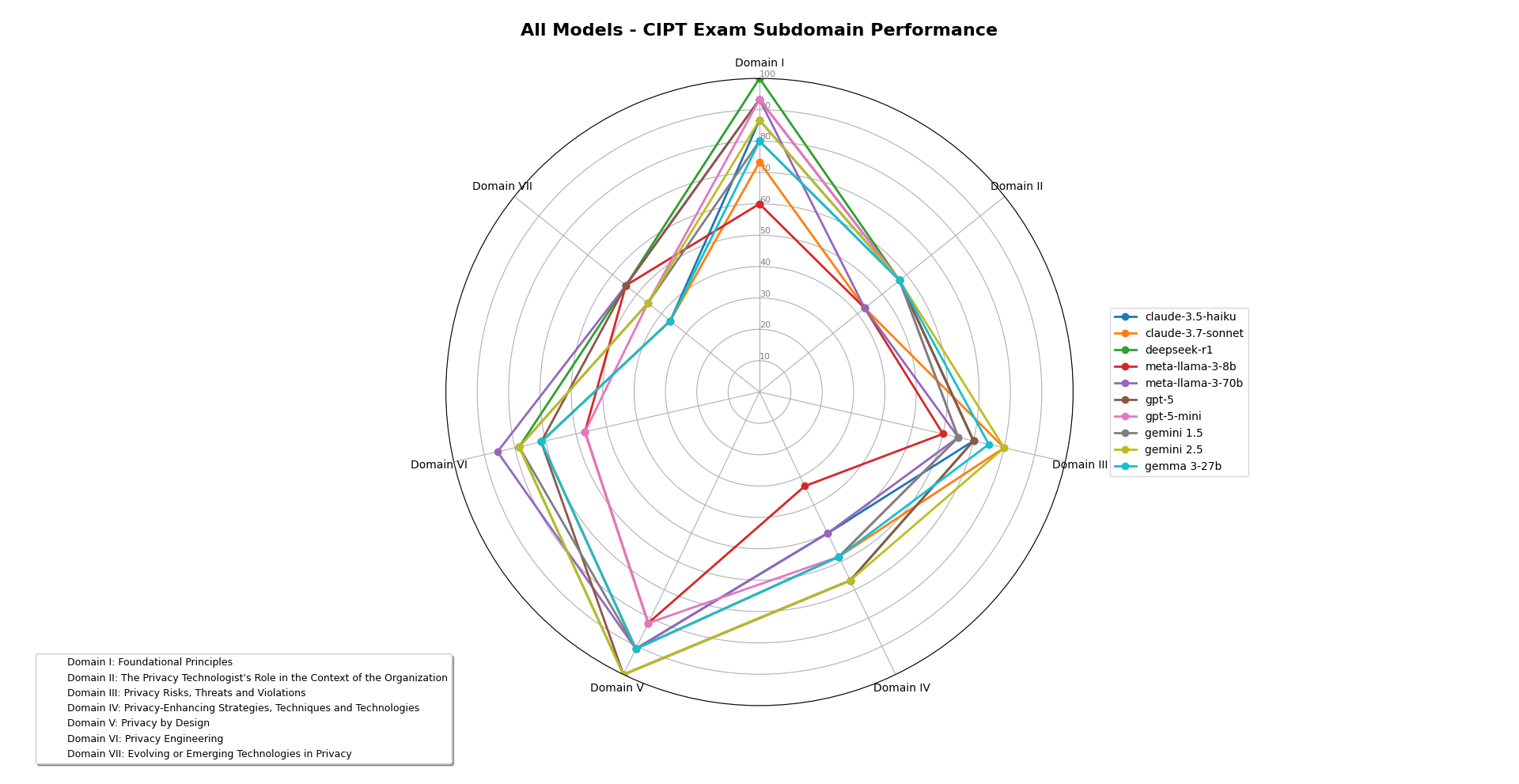}
    \caption{Model performance on the Certified Information Privacy Technologist (CIPT) certification exam by Subdomain}
    \label{fig:enter-label}
\end{figure}

The subdomain analysis of the CIPT exam reveals a clear grouping of consistent strengths and weaknesses across the models. All models generally scored very well on "Foundational Principles" and "Privacy by Design", but no models scored above 66.7\% on “Evolving or Emerging Technologies in Privacy”, "Privacy-Enhancing Strategies, Techniques and Technologies", and "Evolving or Emerging Technologies in Privacy". Other technical areas like “Privacy Engineering” and “Privacy Risks, Threats and Violations” presented less difficulty for advanced models, which scored 70–80\%, while meta-llama-3-8b lagged consistently, never outperforming any of its larger peers in any section.

\subsection{AIGP Model Benchmark Results}
\begin{figure}[H]
    \centering
    \includegraphics[width=1\linewidth]{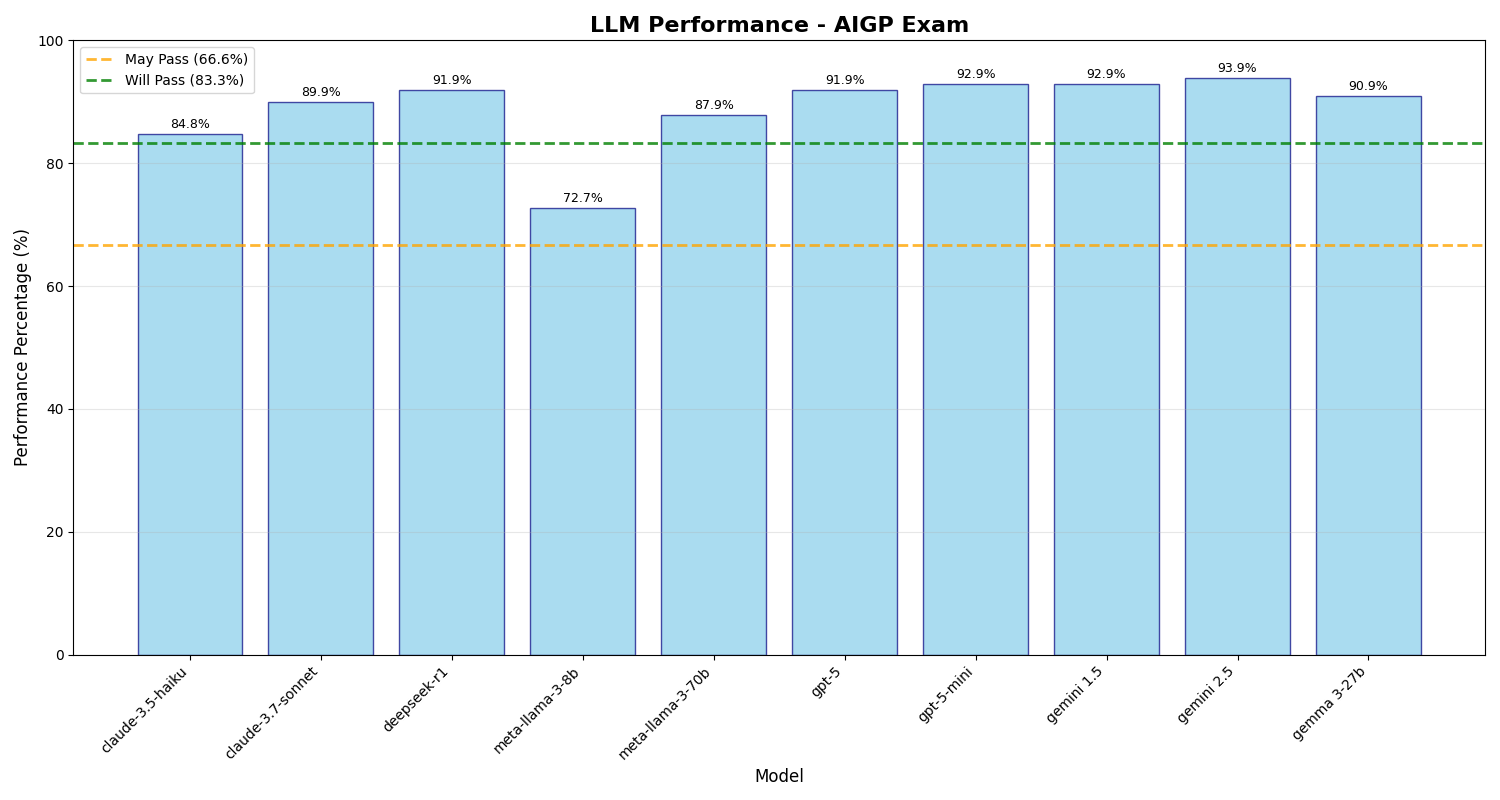}
    \caption{Model performance (percentage of correct answers) on the AIGP (AI Governance Professional) certification exam}
    \label{fig:enter-label}
\end{figure}

The Artificial Intelligence Governance Professional (AIGP) exam saw strong results for every model - with nine out of ten clearing the guaranteed pass threshold and all models performing well enough to likely pass. The underperforming model was once again meta-llama-3-8b at 72.7\%. The best score was 93.9\% for Gemini 2.5. Most other models, including gpt-5, gpt-5-mini, deepseek-r1, and gemma 3-27b, were tightly clustered in the upper 80\% and lower 90\% ranges, comfortably above the 83.3\% “definite pass” line. 

Notably, gpt-5-mini outperformed the more advanced gpt-5 model on the AIGP exam at 92.9\% and 91.9\% respectively. This is the only case where we observed a cost-saving model outperform its premier model counterpart.

The AIGP exam does not have Domains like the other IAPP exams.

\subsection{Aggregate Model Benchmark Results}
\begin{figure} [H]
    \centering
    \includegraphics[width=1\linewidth]{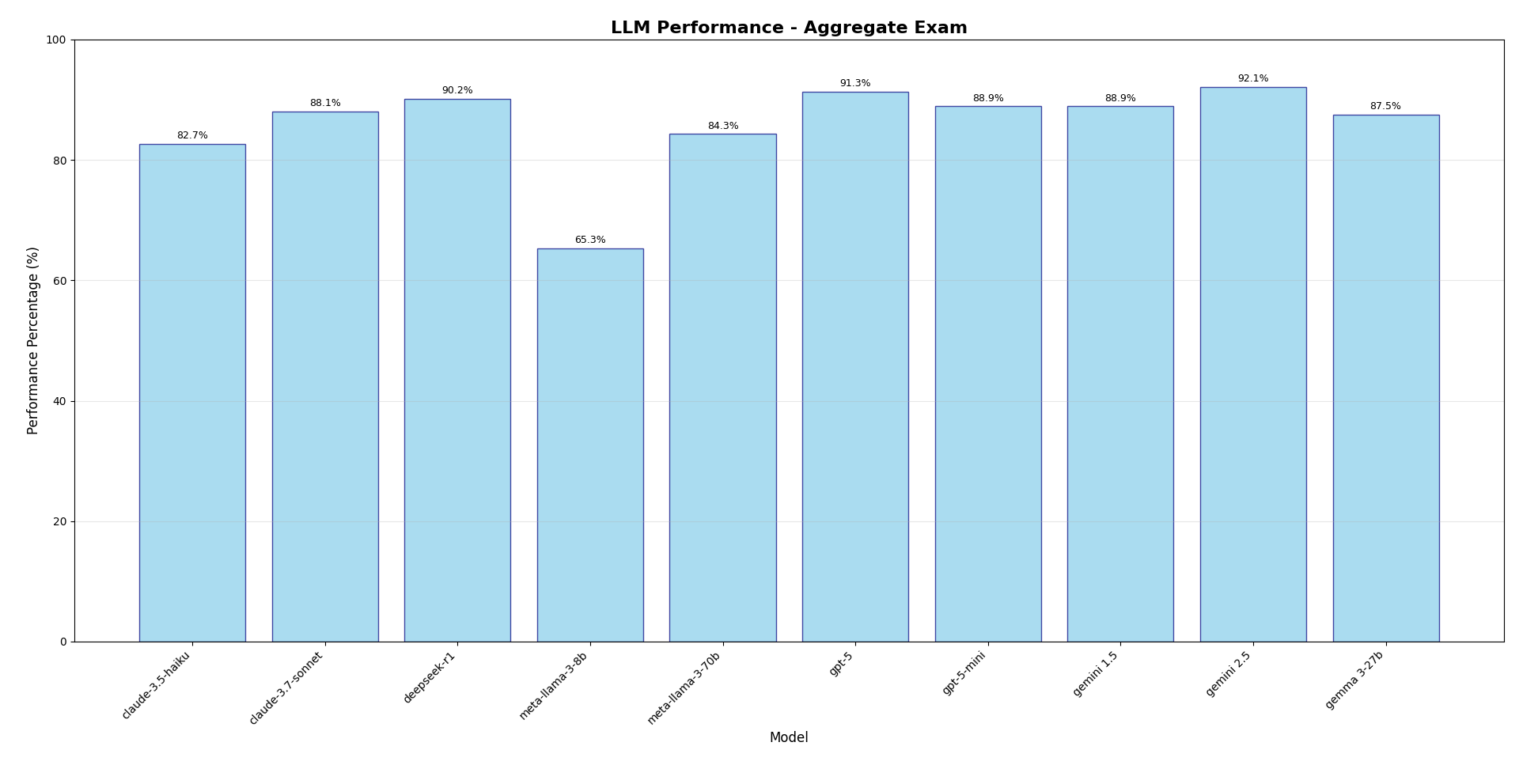}
    \caption{Model performance percentage of correct answers across CIPM, CIPT, CIPP/US, and AIGP exams. }
    \label{fig:enter-label}
\end{figure}

Aggregate performance across all exams shows very strong performance across almost all models. Gemini 2.5 leads narrowly overall with an average score of 92.1\%, closely followed by gpt-5 at 91.3\%. Other strong performers include deepseek-r1 (90.2\%), gemini 1.5 (88.9\%), gpt-5-mini (88.9\%), and gemma 3-27b (87.5\%), all of which would be fantastic performances if delivered by humans. In the middle range, meta-llama-3-70b (84.3\%), and claude-3.5-haiku (82.7\%) display reliable, though slightly less robust, competence. Meta-llama-3-8b, however, lags significantly behind with an aggregate of just 65.3\%, underscoring a pronounced divide in model capability based on scale and architecture.

\begin{figure} [H]
    \centering
    \includegraphics[width=1\linewidth]{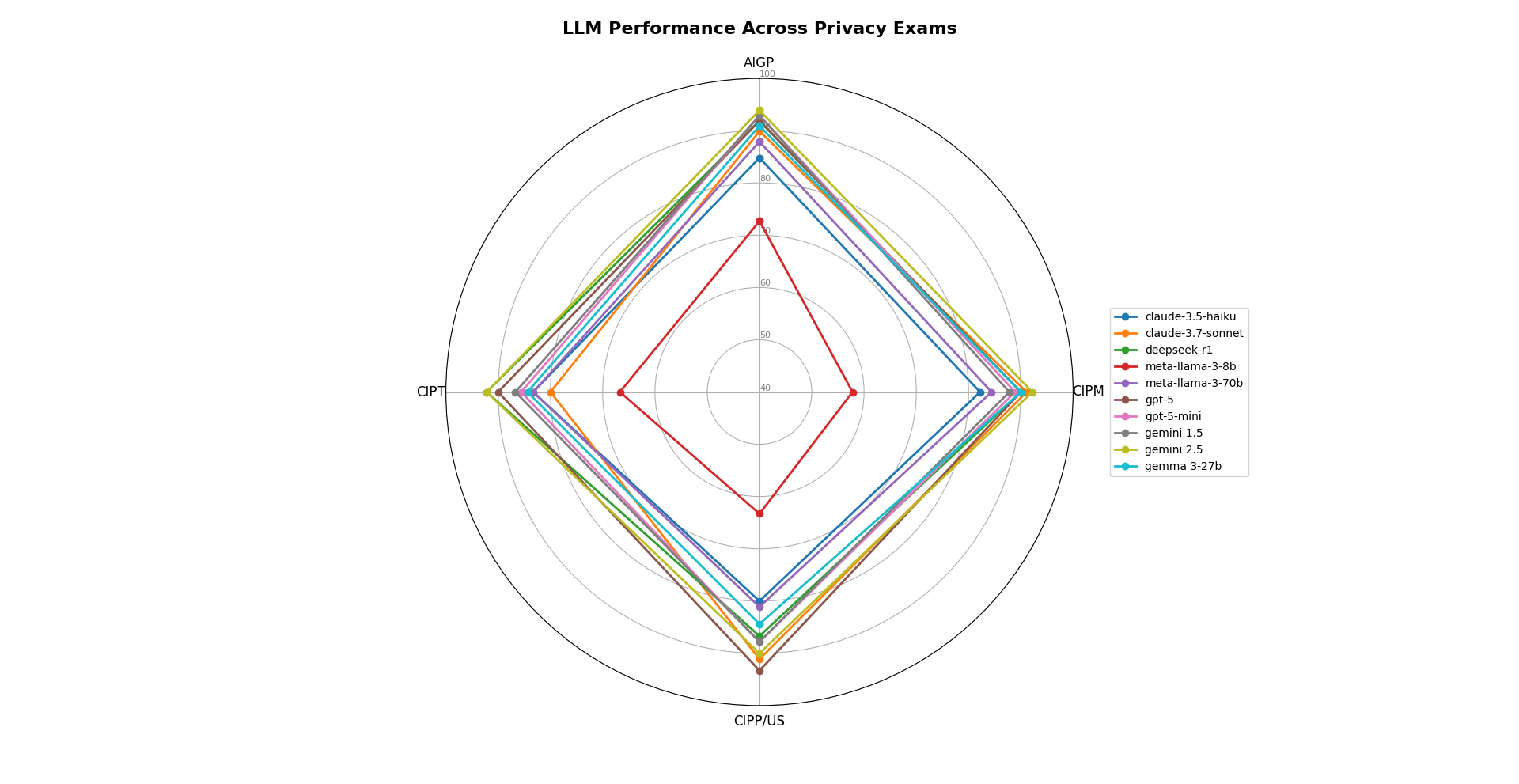}
    \caption{Model performance percentage of correct answers across CIPM, CIPT, CIPP/US, and AIGP exams. }
    \label{fig:enter-label}
\end{figure}

When comparing relative performance across the different exams, models performed best on the AIGP exam with an average score of 89\%, indicating strong comprehension of AI governance, ethics, and risk management topics. Results on the other exams were consistently matched: CIPM averaged 85.4\%, CIPP/US was close behind at 84.6\%, and CIPT followed at 84.4\%. The clustering of scores for CIPM, CIPP/US, and CIPT highlights slightly greater difficulty or less domain familiarity for models on managerial, legal, and technical privacy subjects than on AIGP content. Overall, these results suggest robust and stable model competence across all IAPP exams, with highest confidence and least variability exhibited on the AIGP exam.

\section{Linking Model Strengths to Exam Outcomes}

The benchmarking results reveal clear patterns linking each model’s strengths to their exam performance. Models consistently excelled when test domains matched their core training areas and lagged where coverage was weaker. Gemini 2.5 led globally with an aggregate score of 92.1\%, topping the AIGP (93.9\%) and CIPT (92.2\%) exams and maintaining high marks on both CIPM (92.2\%) and CIPP/US (90.0\%). Deepseek-r1 also demonstrated robust general performance, earning 90.2\% overall, highlighted by 92.2\% on CIPT and 91.9\% on AIGP, and performing solidly on CIPM (90.0\%) and CIPP/US (86.7\%). Models such as Claude 3.7 Sonnet and gpt-5 each maintained high, well-balanced results across every exam - Claude 3.7 Sonnet scoring between 88.1–91.1\% and gpt-5 spanning 90–93.3\%. By contrast, meta-llama-3-70b reached 84.3\% overall but fell below high-proprietary marks, while meta-llama-3-8b lagged across the board at just 65.3\% aggregate and even lower on CIPM (57.8\%) and CIPP/US (63.3\%).

Open models like gemma 3-27b delivered results that closely tracked the proprietary leaders, with 90.0\% on both CIPT and CIPM and 90.9\% on AIGP, illustrating the effect of targeted governance-focused pretraining. In subdomain analysis, these models showed particular strength on exams demanding governance and technical depth. Meanwhile, correlation patterns emerged: AIGP, CIPP/US, and CIPT scores closely tracked one another, indicating strong overlap in legal and technical privacy knowledge. In contrast, CIPM scores - averaging 84.7\% across models - showed weaker correlation with other exams, highlighting that privacy program management content remains a distinct challenge. Most notably, across the field, larger and more generalist models such as Gemini 2.5, deepseek-r1, and Claude 3.7 Sonnet maintained stable, top-level performance everywhere, while smaller models and those lacking domain-specific tuning displayed pronounced peaks and troughs. These findings suggest that while broad pretraining ensures reliable competence, meaningful gaps persist in managerial privacy content - gaps which could be narrowed through case study training or fine-tuning on standards like ISO 27701. In sum, this multi-exam benchmark offers an integrated view of model expertise, pinpointing where broad and specialist approaches converge and where additional focus is most needed.

\begin{figure}[H]
    \centering
    \includegraphics[width=1\linewidth]{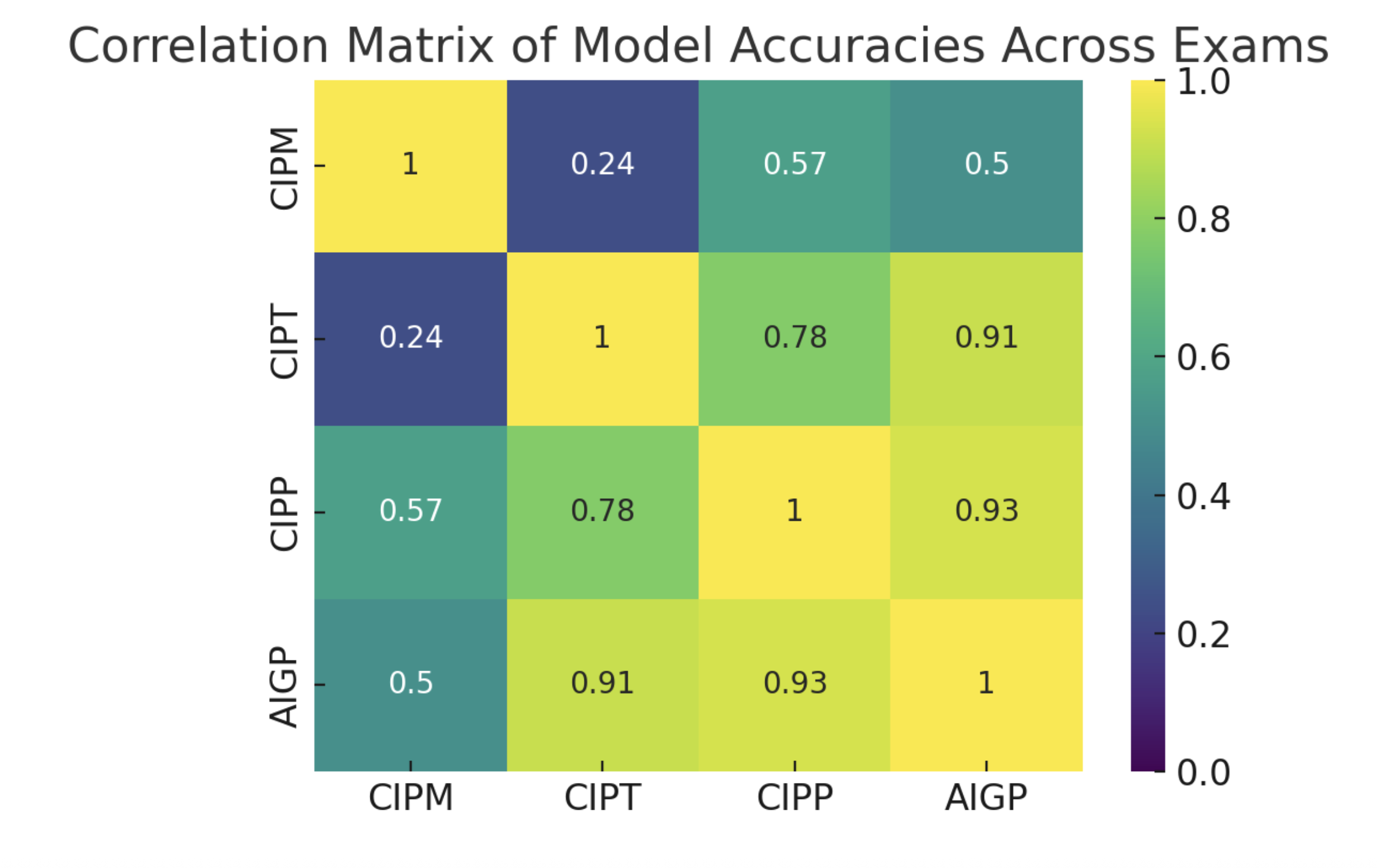}
    \caption{Correlation between Exam Scores}
    \label{fig:enter-label}
\end{figure}

To better understand the relationships in model performance across these exams, we examined the correlation between exam scores. Models that did well on one exam often did well on others, but this was not uniformly true for all exam pairings. For instance, there was a very high correlation between model scores on CIPT and AIGP (Pearson r = 0.91) as well as between CIPP/US and AIGP (r = 0.93). This means that, in general, a model that excelled in technical privacy (CIPT) or privacy law (CIPP) also tended to excel in AI governance – intuitively plausible given that AI governance draws upon both technical understanding and regulatory knowledge. By contrast, the correlation between CIPM and CIPT scores was only about 0.24, which is extremely low. In fact, CIPM scores had the weakest correlation with the other three exams, hinting that the skill set or knowledge required for privacy program management is somewhat distinct. In summary, the raw results indicate that most advanced LLMs can achieve passing or even expert-level scores on all four privacy/governance exams, with the notable exception that certain specialized models might have blind spots as seen with DeepSeek on CIPM. Next, we discuss what these outcomes reveal about each model’s strengths and the nature of the exam domains.

\section{Further Improvements}
 We recognize the significant impact that prompt construction has on the performance of LLM models. Future research could explore strategies to provide more context in the prompt that could dramatically impact the benchmarking performance. The benchmarking could be adapted to accept more complex questions than multiple-choice. Further, we can extend this evaluation to other related certifications or standards – for instance, testing LLMs on privacy laws of other jurisdictions ,for instance, CIPP/E for Europe, or on ethical hacking and see if similar patterns hold. Additionally, as new models emerge or existing ones like Claude and GPT evolve, re-running these benchmarks will track progress over time. We expect that with the rapid pace of AI development, the ceiling of performance may yet rise, and models’ understanding of niche domains will continue to deepen. Ultimately, studies like this contribute to governance of AI – by assessing what AI knows about governance, we can better govern AI’s role in our organizations. 

\section{Conclusion}
This study evaluated ten leading LLMs on privacy and AI governance certification exams. It found that the best models match or exceed human-level performance. Google’s Gemini 2.5, OpenAI’s GPT-5, and Deepseek's R1 scored above 90 percent on all four tests. This indicates that they have a strong corpus of training data sets containing privacy laws such as GDPR and CCPA as well as technical privacy controls and new AI governance principles. Most models scored in or above the passing range for every exam. The open, 27-billion-parameter Gemma-3-27B, which was fine-tuned on governance data, performed similarly to much larger proprietary systems, showing that targeted tuning can build deep expertise.
With only one exception in the AIGP exam, the cost-optimized models did not perform well compared to their flagship LLM counterparts overall. Performance depends not only on the size of the model but also on the range and focus of the training. The weak connection between CIPM scores and the scores on other exams highlights the need for better management content. At the same time, the strong overlap among privacy law, technical privacy, and AI governance suggests that improvements in one area can strengthen the others.

While the exams are not comprehensive of all of the work, judgment, and discrete tasks required to be a successful Privacy or AI Governance professional, the ability of LLMs to pass professional certifications indicates competencies in the foundations of privacy practice. This evidence suggests that with further development and specialization, LLMs have the capability to assist privacy professionals in discrete tasks like drafting privacy policies, answering compliance questions, and conducting automated risk checks. The privacy professional may be able to use LLMs today to significantly augment, automate, or accelerate their work in a significant way. Overall, AI LLMs are passing human benchmarks for complex legal, technical, and organizational scenarios with great accuracy. Still, further research and development is needed to address the remaining gaps in domain knowledge.

\newpage

\appendix
\section{Appendix}
\subsection{Aggregate Exam Scores by Model}

\scriptsize
\begin{table}[ht]
\centering
\begin{adjustbox}{max width=\textwidth}
\begin{tabular}{r l c 
                *{11}{>{\centering\arraybackslash}p{0.9em}} c}
\textbf{\#Q} & \textbf{Exam} & \textbf{Dn} &
\rotatebox{90}{\textbf{C3.5 H}} & 
\rotatebox{90}{\textbf{C3.7 S}} &
\rotatebox{90}{\textbf{DS-r1}} &
\rotatebox{90}{\textbf{L3-8b}} &
\rotatebox{90}{\textbf{L3-70b}} &
\rotatebox{90}{\textbf{gpt-5}} &
\rotatebox{90}{\textbf{gpt-5-m}} &
\rotatebox{90}{\textbf{G1.5}} &
\rotatebox{90}{\textbf{G2.5}} &
\rotatebox{90}{\textbf{G3-27b}} &
\rotatebox{90}{\textbf{Avg.\%}} \\
\hline
99  & AIGP    & All & 84 & 89 & 91 & 72 & 87 & 91 & 92 & 92 & 93 & 90 & 88.4 \\
90  & CIPM    & All & 74 & 82 & 81 & 52 & 76 & 81 & 80 & 79 & 83 & 81 & 84.7 \\
18  & CIPM    & 1   & 15 & 15 & 17 & 12 & 16 & 17 & 16 & 16 & 17 & 16 & 86.7 \\
14  & CIPM    & 2   & 14 & 14 & 14 & 8  & 14 & 13 & 14 & 12 & 14 & 13 & 92.1 \\
20  & CIPM    & 3   & 18 & 18 & 16 & 14 & 16 & 18 & 16 & 16 & 18 & 17 & 83.0 \\
15  & CIPM    & 4   & 11 & 14 & 12 & 8  & 13 & 13 & 14 & 12 & 14 & 12 & 82.7 \\
6   & CIPM    & 5   & 5  & 5  & 5  & 3  & 5  & 5  & 5  & 6  & 5  & 5  & 81.7 \\
17  & CIPM    & 6   & 12 & 16 & 15 & 8  & 14 & 14 & 15 & 14 & 15 & 14 & 81.2 \\
90  & CIPP/US & All & 72 & 82 & 78 & 57 & 73 & 84 & 79 & 79 & 81 & 76 & 84.2 \\
35  & CIPP/US & 1   & 30 & 33 & 30 & 26 & 30 & 33 & 31 & 31 & 31 & 31 & 86.9 \\
25  & CIPP/US & 2   & 16 & 20 & 18 & 16 & 18 & 19 & 19 & 19 & 18 & 18 & 72.4 \\
6   & CIPP/US & 3   & 5  & 6  & 5  & 5  & 5  & 5  & 5  & 6  & 5  & 5  & 88.3 \\
10  & CIPP/US & 4   & 9  & 10 & 10 & 7  & 9  & 10 & 10 & 10 & 10 & 10 & 95.0 \\
14  & CIPP/US & 5   & 9  & 11 & 11 & 5  & 9  & 11 & 11  & 10 & 10 & 8  & 65.7 \\
90  & CIPT    & All & 75 & 72 & 83 & 60 & 75 & 81 & 77 & 78 & 83 & 76 & 83.8 \\
15  & CIPT    & 1   & 13 & 11 & 15 & 9  & 14 & 14 & 14 & 12 & 13 & 12 & 84.0 \\
7   & CIPT    & 2   & 4  & 3  & 4  & 3  & 3  & 4  & 4  & 4  & 4  & 4  & 51.4 \\
20  & CIPT    & 3   & 14 & 16 & 14 & 12 & 13 & 14 & 13 & 13 & 16 & 15 & 71.0 \\
12  & CIPT    & 4   & 6  & 7  & 8  & 4  & 6  & 8  & 7  & 7  & 8  & 7  & 56.7 \\
11  & CIPT    & 5   &10  &10  &11  & 9  &10  &11  &9   &10  &11  &10  &93.6 \\
14  & CIPT    & 6   &10  &10  &11  & 8  &12  &10  &8   &11  &11  &10  &72.9 \\
11  & CIPT    & 7   & 4  & 4  & 6  & 6  & 6  & 6  & 5  & 5  & 5  & 4  &46.4 \\
369 & Aggreg. & All &305 &325 &333 &241 &311 &337 &328 &328 &340 &323 &85.3 \\
\end{tabular}
\end{adjustbox}
\caption{Aggregate exam scores by model and domain (condensed format).}
\end{table}

\subsection{Python Script for Automated Exam Scoring}

The following script was used to automate multiple-choice exam evaluation

The full codebase used for this research is available at: \href{https://github.com/trustsuperset/privacy_llm_benchmark}{\url{https://github.com/trustsuperset/privacy_llm_benchmark}}

\begin{lstlisting}[language=Python, basicstyle=\scriptsize\ttfamily, frame=single, breaklines=true, caption={Python script for evaluating model performance on practice exams.}]
# Install dependencies (if not already installed)
!pip install replicate

# STEP 1: Imports
import os
import pandas as pd
from tqdm import tqdm
import replicate
from getpass import getpass
from google.colab import drive
import re

# STEP 2: Set up Replicate
os.environ["REPLICATE_API_TOKEN"] = getpass(" Enter your Replicate API Token: ")

# STEP 3: Google Drive setup
drive.mount('/content/drive', force_remount=True)
input_folder = "/content/drive/MyDrive/research/practice_exams"
output_folder = "/content/drive/MyDrive/research/replicate_responses"
os.makedirs(output_folder, exist_ok=True)

# STEP 4: Choose Models
models = {
    "meta/meta-llama-3-8b-instruct": "meta-llama-3-8b-instruct",
    # Additional models can be added here
}

# STEP 5: Select Exam Files
exam_files = [
    "CIPT Practice Exam.csv",
    # Add more exam files as needed
]

# STEP 6: Helper Functions
def build_prompt(row, answer_columns):
    lines = []
    if 'scenario' in row and pd.notna(row['scenario']):
        lines.append(f"Context:\n{row['scenario']}\n")
    lines.append(f"Question:\n{row['question']}\n")
    lines.append("Choices:")
    for i, col in enumerate(answer_columns):
        lines.append(f"{chr(65 + i)}. {row[col]}")
    lines.append(
        "\nYou are a certified U.S. privacy professional taking a high-stakes multiple-choice exam.\n"
        "Read the question and choices carefully. Eliminate clearly incorrect choices if possible. "
        "Choose the BEST answer. Respond only in this format:\n"
        "Final Answer: \nExplanation: \n\n"
        "Do not explain all four choices-just support your final choice."
    )
    return "\n".join(lines)

def query_model(model_id, prompt):
    try:
        output = replicate.run(model_id, input={"prompt": prompt})
        return "".join(output) if isinstance(output, list) else output
    except Exception as e:
        print(f" Model query failed: {e}")
        return ""

def extract_by_content_match(response, answer_choices):
    response = response.lower()
    best_match, best_score = "", 0
    for letter, text in answer_choices.items():
        if not isinstance(text, str): continue
        overlap = len(set(text.lower().split()) & set(response.split()))
        if overlap > best_score:
            best_score = overlap
            best_match = letter
    return best_match.upper()

def extract_answer_letter(response, answer_choices=None):
    if not isinstance(response, str): return ""
    match = re.search(r"final answer\s*[:\-]?\s*([A-D])\b", response, re.IGNORECASE)
    if match: return match.group(1).upper()
    lines = response.strip().splitlines()
    if not lines: return ""
    for line in lines:
        if re.fullmatch(r"[A-Da-d]", line.strip()):
            return line.strip().upper()
    match = re.search(r"\b(correct|best|answer)\s*(is|:)?\s*([A-D])\b", response, re.IGNORECASE)
    if match: return match.group(3).upper()
    match = re.match(r"^\s*([A-D])[\.:\-]?\s", lines[0])
    if match: return match.group(1).upper()
    if answer_choices:
        return extract_by_content_match(response, answer_choices)
    return ""

def evaluate_response(model_letter, correct_letter, choices, response):
    model_letter = model_letter.strip().upper()
    correct_letter = correct_letter.strip().upper()
    if model_letter == correct_letter:
        return 1
    correct_text = choices.get(correct_letter, "").strip().lower()
    if correct_text and correct_text in response.lower():
        return 1
    return 0

# STEP 7: Main Process
def process_exam(model_id, model_name, file_name):
    print(f" Processing {file_name} with {model_name}")
    file_path = os.path.join(input_folder, file_name)
    df = pd.read_csv(file_path)
    answer_cols = sorted([col for col in df.columns if col.lower().startswith("answer")])
    required = ['question', 'correct answer'] + answer_cols
    if not all(col in df.columns for col in required):
        raise ValueError(f"Missing required columns in {file_name}")
    optional_cols = ['scenario'] if 'scenario' in df.columns else []
    df = df[required + optional_cols].dropna(subset=['question'])
    results = []
    for _, row in tqdm(df.iterrows(), total=len(df), desc=" Asking model"):
        prompt = build_prompt(row, answer_cols)
        response = query_model(model_id, prompt)
        answer_choices = {chr(65 + i): row[col] for i, col in enumerate(answer_cols)}
        model_letter = extract_answer_letter(response, answer_choices)
        score = evaluate_response(model_letter, row['correct answer'], answer_choices, response)
        result = {
            "question": row['question'],
            "scenario": row.get('scenario', ""),
            "correct answer": row['correct answer'],
            "model answer": model_letter,
            "score": score,
            "response": response,
            **{col: row[col] for col in answer_cols}
        }
        results.append(result)
    df_out = pd.DataFrame(results)
    columns_order = ['question']
    if 'scenario' in df_out.columns:
        columns_order.append('scenario')
    columns_order += answer_cols + ['correct answer', 'model answer', 'score', 'response']
    df_out = df_out[columns_order]
    total_score = df_out['score'].sum()
    total_questions = len(df_out) - 1
    percentage = (total_score / total_questions) * 100 if total_questions > 0 else 0
    score_row = {col: "" for col in df_out.columns}
    score_row['question'] = "TOTAL SCORE"
    score_row['score'] = f"{total_score} / {total_questions} ({percentage:.1f}%)"
    df_out = pd.concat([df_out, pd.DataFrame([score_row])], ignore_index=True)
    output_file = f"{file_name.split('.')[0]}-{model_name}-output.csv"
    df_out.to_csv(os.path.join(output_folder, output_file), index=False)
    print(f" Saved to: {output_file}")
    print(f" Score: {total_score} / {total_questions} ({percentage:.2f}%)")

# STEP 8: Run
for model_id, model_name in models.items():
    for exam_file in exam_files:
        process_exam(model_id, model_name, exam_file)
\end{lstlisting}

\end{document}